# Non-Hermitian topolectrical circuit sensor with high sensitivity

Hao Yuan[1*], Weixuan Zhang[1*], Zilong Zhou[2*], Wenlong Wang[2], Naiqiao Pan[1], Yue Feng[2+], Houjun Sun[3], and Xiangdong Zhang[1$]

[1]Key Laboratory of advanced optoelectronic quantum architecture and measurements of Ministry of Education, Beijing Key Laboratory of Nanophotonics & Ultrafine Optoelectronic Systems, School of Physics, Beijing Institute of Technology, 100081, Beijing, China

[2]School of Mechatronical Engineering, Beijing Institute of Technology, Beijing 100081, China

[3]Beijing Key Laboratory of Millimeter wave and Terahertz Techniques, School of Information and Electronics, Beijing Institute of Technology, Beijing 100081, China

*These authors contributed equally to this work. $+Author to whom any correspondence should be addressed. E-mail: zhangxd@bit.edu.cn; fengyue@bit.edu.cn

**Electronic sensors play important roles in various applications, such as industry and environmental monitoring, biomedical sample ingredient analysis, wireless networks and so on. However, the sensitivity and robustness of current schemes are often limited by the low quality-factors of resonators and fabrication disorders. Hence, exploring new mechanisms of the electronic sensor with a high-level sensitivity and a strong robustness is of great significance. Here, we propose a new way to design electronic sensors with superior performances based on exotic properties of non-Hermitian topological physics. Owing to the extreme boundary-sensitivity of non-Hermitian topological zero modes, the frequency shift induced by boundary perturbations can show an exponential growth trend with respect to the size of non-Hermitian topolectrical circuit sensors. Moreover, such an exponential growth sensitivity is also robust against disorders of circuit elements. Using designed non-Hermitian topolectrical circuit sensors, we further experimentally verify the ultra-sensitive identification of the distance, rotation angle, and liquid level with the designed capacitive devices. Our proposed non-Hermitian topolectrical circuit sensors can possess a wide range of applications in ultra-sensitive environmental monitoring and show an exciting prospect for next-generation sensing technologies.**

Sensing with high precision is of fundamental importance in modern society and technology. There are various schemes for the construction of sensors relying on resonant structures in different physical systems, [1-15] where the shifting and splitting of frequency spectra are always used to identify the external perturbation. For example, the photonic microcavity sensor with an ultra-high quality-factor can be used to monitor the change of background refractive index, and the label-free detection of single molecules can be realized. [1-3] The optomechanical transducer is able to be used as an ultrasensitive detector of weak incoherent forces. [4] In particular, electronic sensors can also offer excellent performances in monitoring multiple environmental parameters. [5-15] There are various types of electronic sensors, such as capacitive sensors, impedance spectroscopy sensors, surface acoustic wave sensors and so on. While, the sensitivity of these circuit sensors is often limited by the low quality-factors of $LC$ resonators, where the magnitude of frequency shifting and splitting are always linearly proportional to the perturbation strength. Moreover, the efficiency and accuracy of existing electronic sensors are also very susceptible to fabrication defects and errors. Hence, exploring new schemes to design electronic sensor with a high-level sensitivity and a strong robustness is of great significance for the next-generation sensing technologies in electronics.

On the other hand, recent advances in the field of topological and non-Hermitian systems have aroused great interest in condensed matter physics and material sciences. [16-26] To date, many classical systems have been designed to fulfill these novel effects in experiments. It is worth noting that different types of disorder-immune topological states could give a variety of new methods for designing topological devices with strong robustness, such as unidirectional waveguides and topological lasers. [27-30] In addition, the ultra-sensitive sensors based on exceptional points (EPs) in non-Hermitian systems have also been proposed and realized in artificial systems, [31-38] where the higher-order root law of frequency splitting of resonators exists under external perturbations. While, EP-based sensors always require a precise tuning of structural parameters, making disorders easily deteriorate the performance of sensors. In this case, it is important to design new types of sensors with strong robustness and high sensitivities.

Recently, engaging studies on the exponential sensitivity with respect to coupling boundaries of the non-reciprocal SSH chain have aroused great interest. [39] Such an exotic lattice model has been experimentally realized based on the mechanical metamaterial, [40] the quantum walks of a single photon [41] and so on. In addition, the developed response theory clearly manifests the strong dependence of non-Hermitian mechanical system on perturbations. [42] Motivated by these studies, the theoretical investigation on the construction of non-Hermitian topological sensor with both a high-level sensitivity and a strong robustness has been proposed, [43] where the sensitivity of sensor grows exponentially with the size of the device. Based on the analytic derivation and numerical calculation, Ref. 43 has pointed out that such a drastic enhancement relies on the anomalous sensitivity to boundary perturbations and is also found to be a stable phenomenon immune to local disorders. It is also shown that the spectral sensitivity of non-Hermitian topological sensor is a suitable quantity to assess the sensing precision in a noise-limited measurement, that is superior to EP sensors. [44, 45] Inspired by these theoretical investigations on the design of non-Hermitian topological sensors, it is important to ask whether the novel electronic sensor with both a high-level sensitivity and a strong robustness can be created.

In this work, we theoretically design and experimentally fabricate an electronic sensor with superior performances relying on exotic properties of non-Hermitian topological physics. With the assistance of high-level sensitivities of non-Hermitian topological zero modes, the boundary perturbation induced frequency shifts of topological zero modes can show an exponential growth trend with the size of non-Hermitian topological sensors being increased. Moreover, such an exponential growth sensitivity is protected by the topological band gap and is also robust against disorders in circuits. Furthermore, we experimentally verify the ultra-sensitive identification of distance, rotation-angle, and liquid-level by non-Hermitian topolectrical circuit sensors. Our work shows an exciting prospect for designing next-generation electronic sensors.

**2. The theory of non-Hermitian topolectrical circuit sensors**

The theoretical model of our designed non-Hermitian topolectrical circuit sensor is schematically shown in **Figure 1a**. There are $N$ nodes in the circuit, and two subnodes in each unit are marked by blue and red dots, respectively. The intercell coupling of the circuit is fulfilled by a capacitor ($C_3$), and the intracell coupling is achieved by an effective non-reciprocal capacitor $C_1 \pm C_2$. The insets of **Figure 1a** display the internal structure of the non-reciprocal capacitor as well as the grounding of two subnodes. The non-reciprocal capacitor is composed of two parts, where one is realized by connecting a capacitor ($2C_2$) in series with a voltage follower and the other one is a capacitor ($C_1 - C_2$) in parallel. In Supporting Information 1, we give the formula derivation of the non-reciprocal capacitor. In addition, different from bulk nodes, which are connected with two adjacent circuit nodes through $C_3$ and $C_1 \pm C_2$, edge nodes are only linked to one nearby circuit node. Hence, to realize the same on-site potential as that of bulk nodes, we need to add an extra grounding capacitor on edge nodes, as plotted in **Figure 1a**. Based on the Kirchhoff equation, the eigen-equation of the designed circuit can be derived.[46-53] The detailed derivation is given in Supporting Information 2. We find that the circuit eigen-equation has an excellent correspondence with that of the non-

Hermitian SSH model. In particular, voltages can be mapped to complex amplitudes of the lattice model. The intercell and intracell couplings are quantized by $C_3/2C_2$ and $(C_1 \pm C_2)/2C_2$, respectively. The eigenenergy of the lattice model ($\varepsilon$) and the eigenfrequency of the electric circuit ($f$) satisfy the relationship of $\varepsilon = f_0^2/f^2 - (C_1 + C_2 + C_3)/2C_2$ with $f_0 = 1/\sqrt{8\pi^2 LC_2}$.

It is well known that the non-Hermitian SSH model has a topologically protected zero-energy mode. To further prove that such a property also exists in our circuit system, we calculate the eigen-spectrum of the designed circuit, as shown in **Figure 1b**. Here, the number of circuit nodes equals to 65, and the value of $C_1$, $C_2$, $C_3$, and $L$ are taken as 1 *nF*, 0.5 *nF*, 1.2 *nF*, and 3.3 *uH*, respectively. It is clearly shown that the frequency region from 1.587 *MHz* to 1.806 *MHz* (marked by the red rectangle) manifests the topological band gap of the designed circuit. The mid-gap eigenmode (marked by the red hexagram) exists at 1.686MHz, corresponding to the topological zero mode. Other eigenmodes marked by black hexagrams are trivial states. Spatial profiles of the topological zero mode and trivial modes in the non-Hermitian topolectrical circuit are calculated, as shown in **Figure 1c and 1d**. It is seen that the topological zero mode and trivial modes are both localized on the boundary of the circuit, but the corresponding physical origins are different. The localization of topological zero mode is intimately tied to the topological property of the designed circuit, where voltage amplitudes at even sublattices is zero protected by the sublattice symmetry. Differently, the boundary-localization of trivial bulk modes is caused by the non-Hermitian skin effect. [54-59] These features are identical to that of the non-Hermitian SSH model.

It is noted that the construction of the non-Hermitian topological sensor with both a high-level sensitivity and a strong robustness can be achieved based on the non-Hermitian SSH model with anomalous boundary sensitivities. [43] In particular, it has been demonstrated that the sensitivity grows exponentially with the size of the device, where the crucial origin is the exponentially decreased overlap of biorthogonal basis for the topological zero mode. In fact, the spectral winding of $k$-space Hamiltonian with periodic boundary conditions could describe the exponential sensitivity on boundary perturbations, where the non-trivial winding number exists with $|C_1 - C_3| < C_2 < |C_1 + C_3|$.[43] Differently, due to the breaking of bulk-boundary correspondence in non-Hermitian systems with skin effects, topological boundary states must be characterized by the spectral winding defined in the generalized Brillouin zone, where the non-zero winding number exists with $C_2^2 + C_3^2 > C_1^2$. The difference of topological origins between these two effects can be explained by the fact that the exponential-sensitive phase transition is related to the degree of localization of the individual left or right eigenvector. In contrast, the phase transition of the topological boundary state is associated with the biorthogonal localization.[43]

To further demonstrate that the high-level sensitivity also exists in our designed electric circuit, we calculate the frequency shifts of topological zero modes by introducing a boundary perturbation to the circuit. The boundary perturbation is achieved by connecting a capacitor $C_s$ (0.1pF), which is much smaller than other capacitors used in the chain, between the first and last nodes. As shown in **Figure 1e** with the red line, the frequency shift of topological zero mode shows an exponential growth trend with the increase of the circuit length, manifesting the exotic sensitivity relating to the non-Hermitian topological zero mode. It is important to note that the maximum frequency shift is limited by the size of the topological band gap. In this case, we can deduce that the frequency shift of topological zero mode induced by an infinite-small boundary perturbation can be amplified to the value limited by the size of the topological band gap with increasing the length of the non-Hermitian topolectrical circuit.

In addition, it is important to show that the exponential-increased sensitivity also possesses a strong robustness protected by the non-trivial topology. To further illustrate the robust sensitivity, we calculate the frequency shift of the topological zero mode in circuits with disorder, which refers to the presence of random fluctuations in all capacitances and inductances. The black, red, green, and blue lines in **Figure 1e** correspond to cases with the disorder strength being 0%, 5%, 15%, and 30%. We can see that the exponential growth trend is still maintained even all circuit elements possess a 30% disorder strength. Furthermore, the slope of frequency shift increases with disorder. This phenomenon is caused by the disorder-induced reduction of the topological band gap.

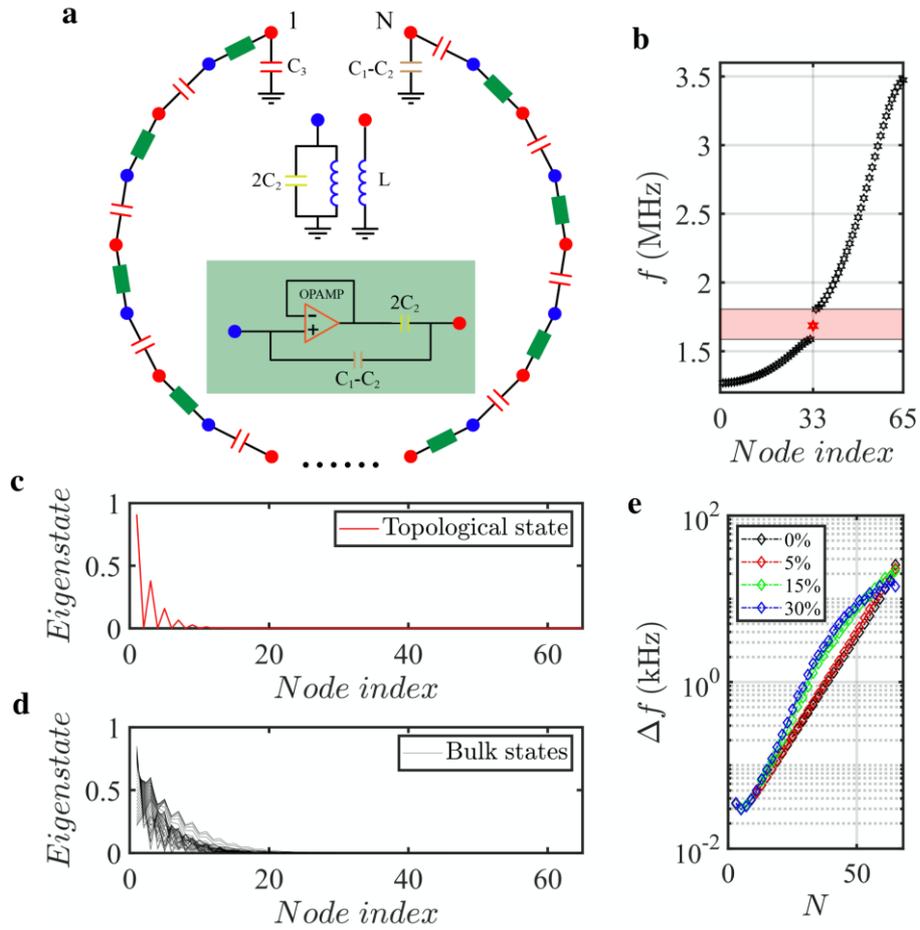

**Figure 1. Theoretical results of the non-Hermitian topolectrical circuit.** a) Schematic diagram of the non-Hermitian topolectrical circuit. Blue and red dots mark two subnodes in each unit. The intracell and intercell couplings are fulfilled by an ordinary capacitor $C_3$ and a non-reciprocal capacitor $C_1 \pm C_2$. Insets present the internal structure of non-reciprocal capacitance and groundings on two subnodes. b)

Sorted eigenfrequencies of the finite non-Hermitian circuit with $N$=65. The red hexagram represents the topological zero mode. c), d) Spatial profiles of the topological zero mode and trivial skin modes in the non-Hermitian topolectrical circuit. e) The frequency shift of topological zero mode as a function of the circuit size under different degrees of disorders.

The detailed analysis and discussions on the influence of disorder are provided in Supporting Information 3. Such a strong robustness exhibits an important advantage compared to the $n$th-order EP sensors, which can only be realized by finely tuning ($2n$-2) parameters. While, our designed non-Hermitian topological circuit does not require the fine-tune of all circuit elements. Hence, we can see that the robust sensitivity of our designed non-Hermitian topolectrical circuit is very beneficial for implementing the ultra-sensitive topological sensing.

**3. Experimental demonstration of the boundary sensitivity for the non-Hermitian topolectrical circuits.**

To experimentally verify that the frequency shift of topological zero mode can exhibit an exponential growth trend with the increase of the circuit length, we fabricate four different-length non-Hermitian topolectrical circuits with $N$=7, $N$=11, $N$=15, and $N$=19. The image of the circuit sample with $N$=19 is shown in **Figure 2a**. The enlarged view of a unit cell and the associated schematic diagram are presented in **Figure 2b**. Three capacitors $C_1 - C_2$ (=0.5 $nF$), $2C_2$ (=1 $nF$) and $C_3$ (=1.2 $nF$), which are used for the intracell and intercell couplings, are enclosed by green, red and blue rectangles, respectively. In addition, the inductor $L$ (=3.3 $uH$) with a high quality-factor is marked by the yellow rectangle. The voltage feedback operational amplifier (Taxes Instrument, LM6171) is utilized to build the voltage follower module. In order to improve the reliability of circuits, some additional capacitances and resistances are introduced. In particular, the resistor $R_1$

(2000Ω) and the capacitor $C_4$ (1000pF) can work as a negative feedback network, which plays a key role in improving the load capacity of the voltage follower and ensures the nearly identical values for the input and output voltages. The resistor $R_2$ (5.1Ω) is used to stabilize the voltage follower and adjust the output current of OpAmp. In addition, two pairs of decoupling capacitors (not shown here) should be added to OpAmp to reduce the power noise. It is worth noting that the first and last circuit nodes are connected by a ceramic capacitor ($Cs$=10pF) to introduce an effective boundary perturbation. In Supporting Information 4, we give a detailed discussion on the selection of circuit elements and the stability analysis of the system.

To characterize the frequency shift of topological zero mode in our designed electric circuits, we measure the impedance responses of the circuit. The red, green, blue, and pink lines in **Figure 2c** present the measured impedance spectra on the first node of different circuits with *N*=7, *N*=11, *N*=15, and *N*=19, respectively. The red block is used to highlight the frequency range of the topological band gap, and midgap impedance peaks correspond to the responses of topological zero modes in circuits. The enlarged view of impedance spectra around the topological zero mode is displayed in **Figure 2d**, where the red dashed line is used to mark the frequency of topological zero mode in the circuit with open boundaries. It is clearly shown that the frequency shift of impedance peaks related to topological zero modes increases significantly with the length of circuits, indicating an enhanced sensitivity with respect to the circuit length. The relationship between measured frequency shifts and circuit lengths is further presented in **Figure 2e** by the blue line. We find that the associated simulation result (illustrated by the red line) is in a good consistence to the experimental result. We can see that the frequency shift of the topological zero mode can show an exponential growth trend with the size of non-Hermitian topolectrical circuits. It is worth noting that the experimental impedance spectra are very stable. Specifically, the amplitudes and frequencies of impedance peaks related to the topological zero-energy modes are nearly unchanged, where the error of frequency shift for the topological zero-energy mode is less than 3 Hz, which is much smaller than the frequency shift of impedance peak. This phenomenon is consistent with the theoretical prediction. In the following, we show that the extreme boundary-sensitivity of our designed non-Hermitian topolectrical circuits can be used to realize the ultra-sensitive identification of different physical quantities.

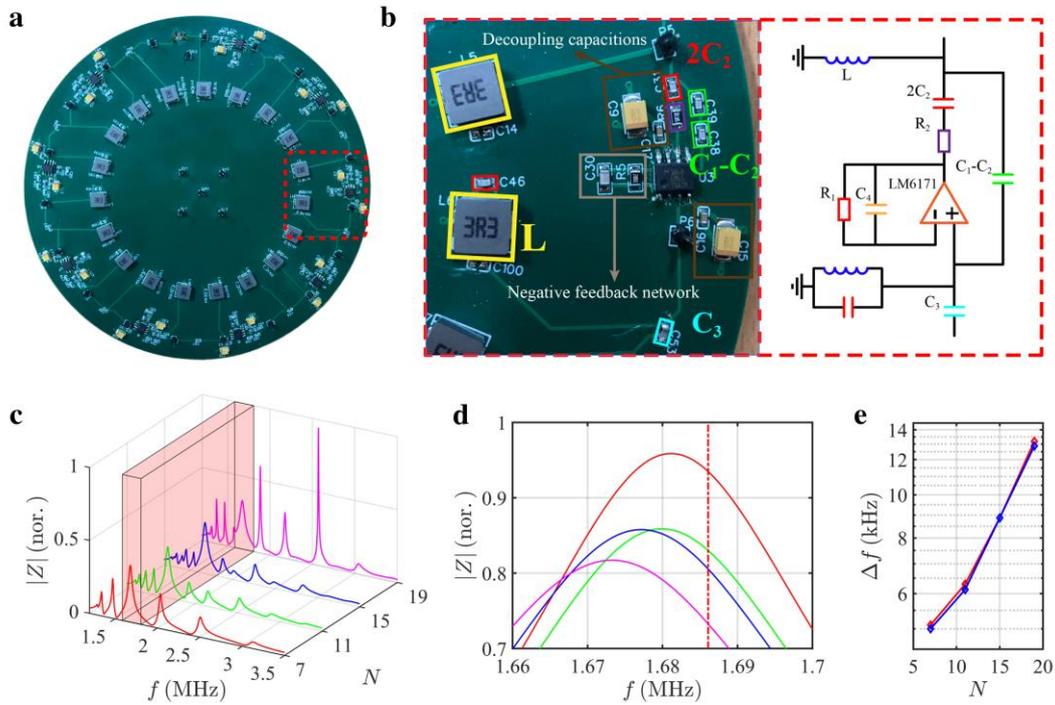

**Figure 2. Experimental results of the non-Hermitian topolectrical circuit.** a) The photograph image of the fabricated circuit with the circuit length being $N=19$. b) The enlarged view of a unit cell and the associated schematic diagram. c) Red, green, blue, and pink lines present the measured impedance spectra on the first node of different circuits with $N=7$, $N=11$, $N=15$, and $N=19$. The topological band gap is marked by the red rectangle. d) The enlarged impedance spectrum around the topological zero mode. The red dashed line is used to represent the frequency of topological zero mode in circuit with open boundaries. e) The blue and red lines illustrate relationships between measured and simulated frequency shifts and the circuit length.

**4. Ultra-sensitive identifications of physical quantities by the non-Hermitian topolectrical circuit sensor**

In this part, we illustrate the ultra-sensitive identification of different physical quantities, including the displacement, rotation angle and liquid level, by our designed non-Hermitian topolectrical circuit sensor. To transfer changes of these physical quantities to the boundary-connected circuit capacitance, three capacitive devices are fabricated. Photograph images of the fabricated displacement, rotation-angle, and liquid-level capacitors are presented in **Figure 3a, 3d, and 3g**. In particular, we find that the variations of measured capacitances of these capacitive devices are linearly related to the change of displacement, rotation angle and liquid (water) level, as shown in **Figure 3b, 3e and 3h**. Detailed design methods and properties of these capacitive devices are provided in Supporting Information 5.

At first, we focus on the identification of displacement by our designed non-Hermitian topolectrical circuit sensor. Here, the change of distance is achieved by tuning the transverse motion of 'electrode 2' in the displacement capacitor (shown in **Figure 3a**). As shown in **Figure 3c**, we measure the frequency shift of the topological zero mode induced by the change of displacement in the displacement capacitor, which is used to connect the first and last nodes in the topolectrical circuit. The red and green diamonds correspond to results of the non-Hermitian topolectrical circuits with $N$=19 and $N$=11, respectively. And, the blue diamond illustrates the result of the Hermitian SSH topological circuit with $N$=19, which possesses an identical eigen-spectrum with that of the non-Hermitian counterpart (see Supporting Information 6 for details). It is clearly shown that the frequency shift of topological zero mode in the non-Hermitian topolectrical circuit with $N$=19 is much larger than that of the non-Hermitian topolectrical circuit with $N$=11 and the Hermitian counterpart. In particular, the ratio of sensitivities of these three topological circuits is about 108: 50: 1, indicating a significant increase of sensitivity by the non-Hermitian topolectrical circuit with a large length. It is worth noting that the relationship between the shift of zero-energy frequency and the strength of boundary perturbation can be explained by perturbation theory. [43] Detailed calculations and discussions are provided in Supporting Information 7.

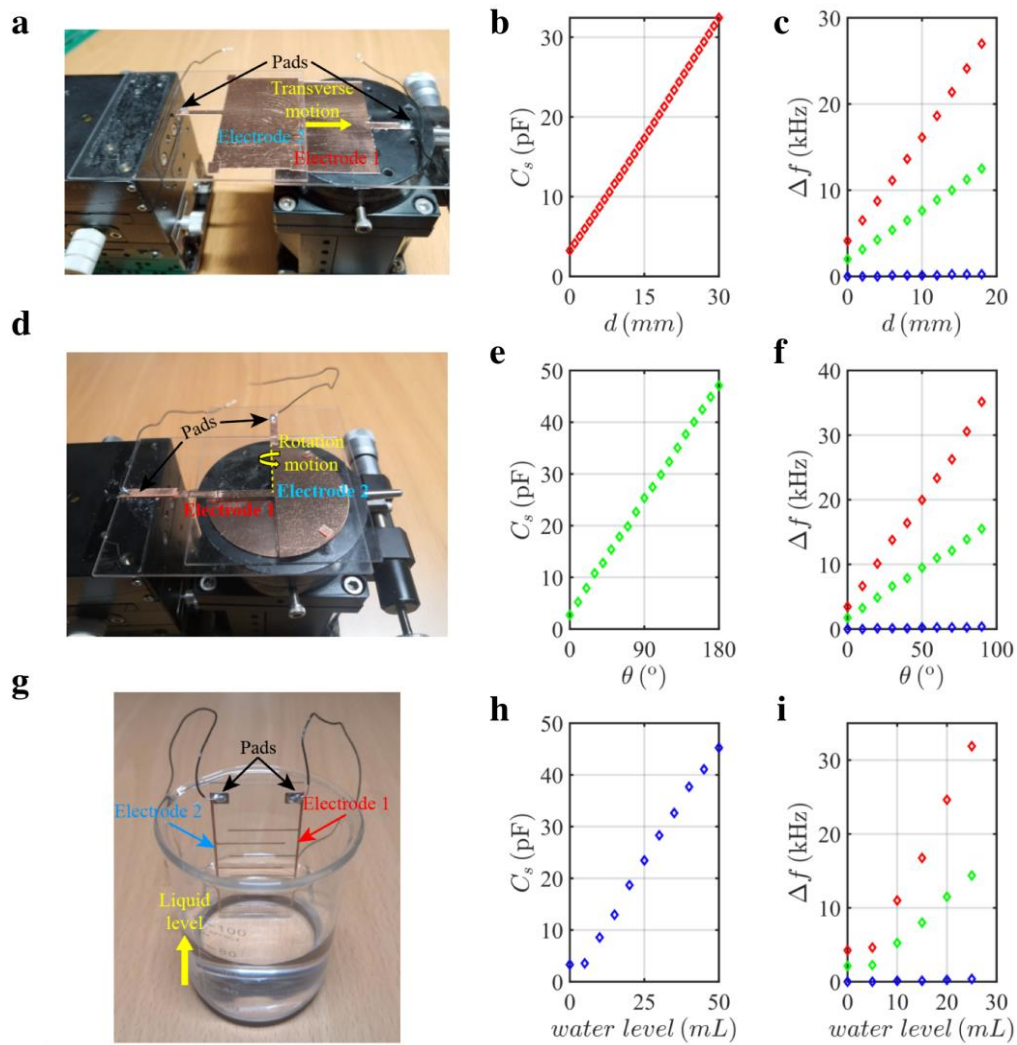

**Figure 3. The ultra-sensitive identifications of displacement, angle and liquid level by the non-Hermitian topolectrical circuit sensor. a), d), g)** Photograph images of the displacement, angle, and liquid-level capacitive devices. **b), e), h)** Relationships between changes of displacement, angle, and liquid level with respect to capacitances of three capacitors. **c), f), i)** The frequency shifts of the topological zero mode induced by the change of displacement, angle and liquid level. Red, blue, and green diamonds correspond to results of the non-Hermitian topolectrical circuits with $N$=19, $N$=11, and

the Hermitian SSH topological circuit with *N*=19, respectively.

Similarly, as shown in **Figure 3f and 3i**, we further measure the frequency shifts induced by the change of rotation angle and water level by connecting the first and last nodes with the designed rotation-angle capacitor and liquid-level capacitor. Specifically, the change of the rotation angle (water level) is realized by adjusting the rotation motion of the 'electrode 2' in the rotation-angle capacitor (the volume of water in the liquid-level capacitor). We can see that the sensitivity of non-Hermitian topolectrical circuit with *N*=19 is also much larger than that of other two topological circuits. These results clearly prove the effectiveness of our designed non-Hermitian topolectrical circuit sensor for realizing the ultra-sensitive identifications of different physical quantities.

**5. Discussion and conclusion**

In conclusion, we theoretically design and experimentally fabricate the non-Hermitian topological circuit sensor with superior performances. In terms of sensitivity, the frequency shift of topological zero mode caused by the boundary perturbation can increase exponentially with the size of the circuit length. In addition, the exponential increase of the sensitivity with respect to the circuit length does not depend on any fine-tuning protected by the topological bandgap. Thus, by using our designed non-Hermitian topolectrical circuit sensor, we experimentally verify the ultra-sensitive identification of distance, rotation angle, and liquid level. It is worth noting that the exponential sensitivity can also exist in the system with even numbers of lattice sites. In addition, the exceptional perturbation in lattice modes with even sites is different from odd-site systems. We note that there are two modes in the topological bandgap for the even-site lattice, which may influence the precision of detected quantities. See detailed analysis and discussions in Supporting Information 8. Furthermore, except for the one-dimensional non-Hermitian topological circuits, we expect that the non-Hermitian topological systems in high dimensions may possess much higher sensitivities compared to low-dimensional counterparts, giving new prospects for the exotic sensing technology.

On a related note, previous studies have pointed out that the influence of noises on the designed non-Hermitian EP-based sensors imposes the fundamental bound on the sensitivity. In this case, the resolvability of the impedance peak for topological zero modes under the influence of noises should be clarified in our designed non-Hermitian topolectrical circuit sensors. It is noted that there are shot noise, flicker noise, and thermal noise in an electronic circuit. Here, we only focus on the thermal noise. Because, the shot noise mainly stems from diode or vacuum tube, and the flicker noise can be eliminated below the level of thermal noise with appropriate resistors.[60] Based on numerical simulations, we find that the voltage noises are increased with respect to the circuit length. This noise effect generates a theoretical limitation on the maximum length of non-Hermitian topolectrical circuit sensors with available performances. In this case, we note that there is an optimal length for our designed electric circuits possessing high-level sensitivity and resolvability at the same time. Additionally, it is important to note that the impedance analyzer possessing a lock-in amplifier can significantly suppress voltage noises, making the experimental impedance noise become much lower than that evaluated in theoretical analysis. The detailed analysis and discussions on the influence of noises on the non-Hermitian topolectrical circuit sensors are provided in Supporting Information 9.

Finally, it is worth noting that our research also opens the possibility of designing non-Hermitian topolectrical circuit sensors at chip scale. There are many advantages on implementing non-Hermitian topolectrical circuit sensors on a chip. In terms of degrees of freedom, all components in a chip are fully configurable. In this case, we can optimize each component according to the requirement of the device to achieve a more stable output of impedance responses. Moreover, we can also use the switching array to achieve the precise regulation of parameters used in experiments, such as capacitances, inductances, OpAmp gains, and bandwidths. In terms of the operation frequency domain, the working frequency of topological zero mode in circuit can be amplified with thousands of times by using chip-scale capacitances and inductances, whose values are much smaller than those used in PCBs. In terms of the sensing capability, it is difficult to accurately identify the weak coupling in a PCB platform with the fore-end capacitance being the same order with the parasitic capacitance in PCBs. Therefore, the reduction of parasitic effect is important in the design of ultrasensitive sensors. Considering the short distances between circuit elements in the chip and the extremely weak parasitic effects, the non-Hermitian topolectrical circuit integrated into a chip can have a superior ability to recognize weak physical quantities. Lastly, in terms of the influence of noises, the effect of noise on the impedance spectrum can be predicted in the chip, since all circuit devices can be precisely formulated. So in these regards, our suggested non-Hermitian topolectrical circuit sensors represent a promising method for designing the ultra-sensitive sensor in the chip.

**Experimental Methods**

*Sample Fabrications*: In the experiment, a voltage feedback operational amplifier (Taxes Instrument, LM6171) is utilized to build the voltage follower module, which blocks the input current while maintaining the output voltage stable. All ceramic capacitors are made from C0G/NP0 material, which ensures the stability of capacitance. To ensure the tolerance of circuit elements and series resistance of inductors to be as low as possible, we use a WK6500B impedance analyzer to select circuit elements with high accuracy (the disorder strength is only 1%) and low losses. In addition, a pair of 2.2uF tantalum capacitors and 1uF ceramic capacitors are used in parallel at both ends of the power supply for all operational amplifiers of all PCBs to realize the function of filtering. We exploit electric circuits by using LCEDA program software, where the PCB composition, stack-up layout, internal layer, and grounding design are suitably engineered. In the PCB design of the short chain circuit structure ($N$=7, 11), we used a double-layer board structure, and in the PCB design of the long chain circuit structure ($N$=15, 19), we used a 6-layer board structure. In the long chain circuit structure with six layers of boards, except for the top and bottom layers, two internal planes in the PCB are used to supply power to all OpAmps, and a grounding signal layer is placed in the middle of the internal planes.

To fabricate capacitive transducers, a fiber laser marking machine (Jinan Consure Electronic Technology Co. Ltd, CS-F30) was used for engraving electrode patterns on single-side-copper clad silica glass sheets with 35 μm thickness of copper conductive layer and 1.0 mm thickness of transparent glass substrate. After that, the sheets were washed with purified ethanol to eliminate the residual contaminants. Then poly(chloro-p-xylylene) films (parylene C, SCS Coatings) with a thickness of 15 μm were deposited on the electrode's surface as the insulation coating and finished with a drying process at 120°C for 5 min. Using the LCR digital bridge (Tonghui, TH2840B) at 1.686 MHz, the capacitances of sensors in response to physical quantities such as the displacement, angle, and liquid-level are obtained.

*Circuit Measurements*: In the measurement of impedance spectra, we connect the impedance analyzer WK6500B interfaces to the GND network of the circuit and a circuit node. Two DC powers (±5V) supply all OpAmps in the PCB.

*Circuit Simulation*: Texas Instruments official Spice model of the OpAmp is used in the circuit simulation, and LTspice is used for numerical simulations.

**Supporting Information**

Supporting Information is available from the Wiley Online Library or from the author.

**Acknowledgements**.

Hao Yuan, Weixuan Zhang, Zilong Zhou contributed equally to this work. This work is supported by the National Key R & D Program of China under No. 2022YFA1404900, National Science Foundation for Young Scientists of China under No. 12104041 and No. 52077005, and the BIT Research and Innovation Promoting Project under No. 2022YCXZ027.